\documentclass[11pt, a4paper]{article}
\usepackage[utf8]{inputenc}
\usepackage[margin=33mm]{geometry}
\usepackage{graphicx}
\usepackage{authblk}
\usepackage{amsmath}

\title{Uncovering structural diversity in commuting networks. Global and local entropy.}
\author[1,*]{Valentina Marin}
\author[1]{Carlos Molinero}
\author[1]{Elsa Arcaute}
\affil[1]{Centre for Advanced Spatial Analysis (CASA) University College London, UK}
\affil[*]{corresponding author: ucqbvpm@ucl.ac.uk}
\date{}

\begin{document}

\maketitle

\begin{abstract}
In this paper we revisit the concept of mobility entropy. Over time, the structure of spatial interactions among urban centres tends to become more complex and evolves from centralised models to more scattered origin and destination patterns. Entropy measures can be used to explore this complexity, and to quantify the degree of structural diversity of in- and out-flows at different scales and across the system. We use toy models of commuting networks to examine global and local measures, allowing the comparison to occur between different parts of the system. We show that entropy at the link and node level give different insights on the characteristics of the systems, enabling us to identify employment hubs and interdependencies between and within different parts of the system. We discuss how these can be used to inform planning and policy decisions oriented towards decentralisation and resilience. 
\end{abstract}

\section{Introduction}

Cities are essentially relational, they are defined by the nature of interactions that holds them together. In the same way, they could also be defined by how they are connected to other cities within a system of intricate relationships. Different types of interactions lead to relationships of dominance, dependency or cooperation between cities, and in doing so, they characterise the functioning and dynamics of the whole system. Systems of cities are interdependent, a significant change in one of its components could impact or disrupt the functioning of other urban entities within the system, or even the structure of the system as a whole.

In this context, systems of cities can be modelled as networks, where different types of links represent different interdependencies, giving rise to different structures \cite{Pumain2006,Bretagnolle2009,batty2018inventing}. Within a myriad of relationships that link urban systems, the connection between workplaces and home has been a central part of studies for understanding the dynamics occurring within systems of cities. Mobility patterns have been widely studied to examine the structure of mobility and its relation with socio-demographic variables \cite{Montis2007,Lenormand2015}, to define categories of cities according to their commuting structure \cite{ Louail2015}, to investigate the evolution of mobility patterns over time \cite{ Patuelli2007,Louail2014}, define boundaries of functional areas \cite{Kropp2014}, and study the spread of infectious diseases \cite{Balcan2009}, to name a few.

When commuting is seen as a network, cities are represented as nodes, and the flows of commuters constitute the links. The structure and characteristics of such a network can give us insights into the different roles that cities play within the system. In particular, diversity and dispersion of flows across the system can inform about the cohesiveness and balance of the relative importance of urban centres. On the other hand, the over concentration of flows can reveal subordination and high dependence of the system in few specific centres, exhibiting the potential susceptibility of the system as a whole. For example, the extent to which in-commuting is concentrated seems crucial to understand potential labour market centralisation and disparities in the distribution of job opportunities. 

Diversity of patterns of labour supply and demand across the territory, is a critical attribute for the resilience of the commuting network. Diverse mobility patterns contribute to the re-organisational capacity of the system \cite{Reggiani2010}. A diverse system has multiple responses and alternatives of meeting a given need, making space for adaptation and innovation to maintain the functioning of the system across different conditions and change \cite{Levinetal1998, berkes2008navigating, Ahern2011, Cumming2013, Marcus2014}. In this context, the relationship between diversity and complexity has gained great attention, particularly from resilience theory \cite{Holling2001UnderstandingSystems}, being a central matter in many fields of science. However widely used, and despite the noted relevance of diversity as a crucial characteristic of urban systems \cite{JJacobs1961, Batty2004, Ahern2011, Bettencourt2014}, it remains a difficult concept to define and measure. The difficulty lies in the many different methodological approaches to measure it across disciplines, encountering great semantic variation \cite{Hamilton2005, Jost2006}.

Entropy is one of the most common ways of quantifying diversity \cite{Page2010}. The concept of entropy was first coined in thermodynamics, and then widely used in other fields such as physics, statistics, information theory and ecology. Depending on the research context, entropy is generally addressed as a measure of disorder in a system, or as the level of uncertainty and information \cite{ben2008farewell}. The latter was introduced by Shannon (1948) in the context of information theory, referring to the amount of information within probability distributions \cite{Shannon1948}. Shannon's entropy measures the degree of uncertainty in predicting the types of elements randomly chosen from a sample. It depends on both, the number of types and the relative abundance of them, also known in the field of ecology as richness and evenness respectively. The greater the amount of types (richness) and the more equally abundant they are (evenness), the more difficult it is to predict \cite{Hamilton2005,heip1998indices}. In such a way, when applied as a diversity measure, one can say that the more uncertainty, the greater the diversity. 

The entropy of network-based systems refers to the heterogeneity in the arrangement of its components. Entropy measures on graphs were first used by Rashevsky (1955) and Mowshowitz (1968) as a measure of relative complexity. First approaches studied the topological information content in unweighted and undirected graphs \cite{Rashevsky1955,Mowshowitz1968}. A common way to measure entropy in graphs is based on the degree distribution, measuring the probability of having a node with a certain number of links. It is a local measure with a focus on node's connectivity which ignores to some degree the weight of the different links. Although it is a useful measure to characterise important aspects of the network, on its own it is unable to describe the complexity of the network structure, from both local and global perspectives. More recent studies have deepened the understanding of entropy in weighted and directed graphs by extending information theory concepts to networks \cite{Sole2004, Wilhelm2007}. This is crucial for studying diversity of commuting networks, which are described by in-commuting and out-commuting flows.

In commuting networks, entropy is commonly used as a relative measure of the distribution of commuters amongst employment locations. This is achieved by looking at the relative abundance and volume of flows embodied in the weights of the links. Although the study of flows by means of entropy has not been widely adopted, certain studies have validated its use for addressing key urban matters such as the analysis of patterns of spatial dispersion to inform choice models for urban transportation \cite{Lowe1998}; the use of in- and out-commuting entropy on different cities to explain variations in economic growth \cite{Goetz2010}; the use of entropy of individual users trajectories to study the correspondence of mobility diversity to social behaviour and socio-economic indicators \cite{Pappalardo2016, Cottineau2019}, as a measure of spatial inequality and attractiveness \cite{Lenormand2020}, or the use of entropy of individual vehicular mobility to characterise spatio-temporal patterns of activities along the day \cite{Gallotti2013}.   

Typically, to the best of our knowledge, measures are carried out at a local level, that is, entropy is calculated for individual trajectories or specific nodes within the network. Cities or administrative areas are the local units of analysis, and there is no wider consideration of the global structure/context to understand their role within the overall regional or national system. On this basis, we think that certain properties of the global network in terms of its structural diversity may have been left unexplored by focusing on the performance of local elements. 

In this paper we test different measures of entropy on commuting networks at global and local scale, considering the probability distribution of flows in both nodes and links. We aim to explore if the results offered by the different measures are complementary and relevant for the study of the structural diversity of spatial interactions. We use toy models of networks with different patterns to examine the different measures and compare the outcomes across systems and their constituent parts. First, we look at the diversity of the global commuting structure by applying a set of measures to both, the group of all nodes and the ensemble of links. We look at nodes based on the workforce patterns of in- and out-commuting flows in all urban units as a result of the spatial distribution of labour supply and demand. The measure depicts patterns of centralisation and dispersion in the urban spatial structure. When studying the network from the perspective of its links, we focus on the diversity of distribution of origin-destination pairs, considering intensity and density of flows. Normalisation of the measure is achieved by comparing the entropy of the sampled links with the maximum possible links of a fully connected network. This measure describes the level of dispersion of commuter trips in the territory, outlining potential functional dependencies when many trips take place in few dominant Origin-Destination (OD) pairs. In a following section we address local entropy at nodes individually. We look at the structural diversity of the sub-graph made up of the subset of interactions that a given urban unit establishes directly with its neighbours. Then we compare the results from the general equation of nodal entropy with a normalised measure that considers its maximum potential if connected to all other nodes in the network. Finally, we discuss the relevance of the twofold analysis of the system comparing global and local approaches, guided by the following questions: Is it the same to be a non-diverse unit within a structural dispersed system, than being a non-diverse unit within a structural concentrated one? To what extend the diversity of the individual elements could describe the diversity of the overall system? What can we learn about the system by comparing the outcomes of local and global scales?

\section{Methods}
The entropy measures presented in this paper are based on the information-theoretic approach to networks \cite{Sole2004, Wilhelm2007}. This approach considers entropy as a measure of uncertainty related to the information content transmitted from sender to receiver. When applied to networks, an analogy is made so effluxes of nodes correspond to the sender and influxes of nodes to the receiver \cite{Wilhelm2007}. Then, the uncertainty of the transmission of a certain flux in the network depends on the probability of its occurrence between sender and receiver. In our case study, commuting networks are constituted by origin and destination nodes representing urban units with in- and out-flows. Measures of uncertainty in Information Theory derive from the Shannon Entropy $H$ formula \cite{Shannon1948}, (also known as Shannon's diversity index in ecology) which is defined as:

\begin{equation} \label{eq:1}
H= -\sum_{\forall i} p_i \log p_i 
\end{equation}
where $N$ is the number of types and $p_{i}$ is the probability of occurrence of the $i$th type within the total sample. 

We use different toy models of commuting networks to compare different measures on both local and global scales applied to the links and nodes of the network. Different forms of normalisation are presented in each case. The following measures consider the commuting flows as directed and weighted graphs $(G)$ represented by a set of $n$ vertices $V(G)$ and $m$ edges $E(G)$, each representing a tuple of nodes. Each node attracts in-commuting flows, and releases out-commuting flows in different proportions depending on its role within the system. For every edge, a weight $w_{ij}$ is assigned, representing the total flow from origin $i$ to destination $j$.

\section{Results}

\subsection{Global diversity}

Global measures quantifying diversity as a function of the overall structure of the commuting network are computed across the whole graph. To measure entropy globally, we look at how flows are distributed either on nodes or on links, considering every component of the network. Given that all elements are interdependent in the overall structure, any local change in the commuting network will modify the global entropy.

\subsubsection*{Spatial distribution of labour supply and demand}
 
Labour supply and demand are not evenly distributed in the geographic space, giving rise to complex patterns of spatial interactions which are reflected in the structure of the commuting network.

Urban units have different functional roles within the system. Some cities for example function as employment hubs, attracting large numbers of workers, other cities mostly supply workers to other areas, while some others are able to find a balance between labour supply and demand. Entropy measures enable us to explore whether the flows in a system tend to be concentrated in dominant areas, or evenly dispersed from many origins to many destinations. The former is characterised by a monocentric pattern where the flows come from many origins to very few destinations, and the latter is characterised by a more polycentric pattern, which indicates a greater balance in the importance of urban units.

Identifying monocentricity or polycentricity through commuting patterns can provide an initial insight into diagnosing possible functional dependencies due to disproportionate concentrations in some central cities. These can serve to inform planning to overcome spatial disparities, with interventions related to labour decentralisation and transport infrastructure aiming at encouraging growth of subordinate areas and more balanced and diverse spatial interaction patterns.

Let us start by characterising origins and destinations through the diversity of locations from/to which workers go/arrive to work. This can be captured through the following global entropy measures:

\begin{itemize}
\item Global out-flow entropy at node level:

\begin{equation}
 H_{GN}^{out}= -\sum_{\forall i}\left(\sum_{\forall j} p_{ij}\right)  \log \left(\sum_{\forall j} p_{ij}\right)
\end{equation} 
where $\sum_{j} p_{ij}$ is the probability of out-flow from node $v_{i}$, considering the sum of all flows departing from $v_{i}$ to every possible node $v_{j}$.

\item Global in-flow entropy at node level:

\begin{equation}
  H_{GN}^{in}= -\sum_{\forall j} \left(\sum_{\forall i} p_{ij}\right)  \log  \left(\sum_{\forall i} p_{ij}\right)
\end{equation} 
where $\sum_{i} p_{ij}$ is the probability of in-flow to node $v_{j}$, considering the sum of all flows arriving to $v_{j}$ from every possible node $v_{i}$.
\end{itemize}

\begin{figure}[h!]
\centering
\makebox[\textwidth][c]{\includegraphics[width=1.15\textwidth]{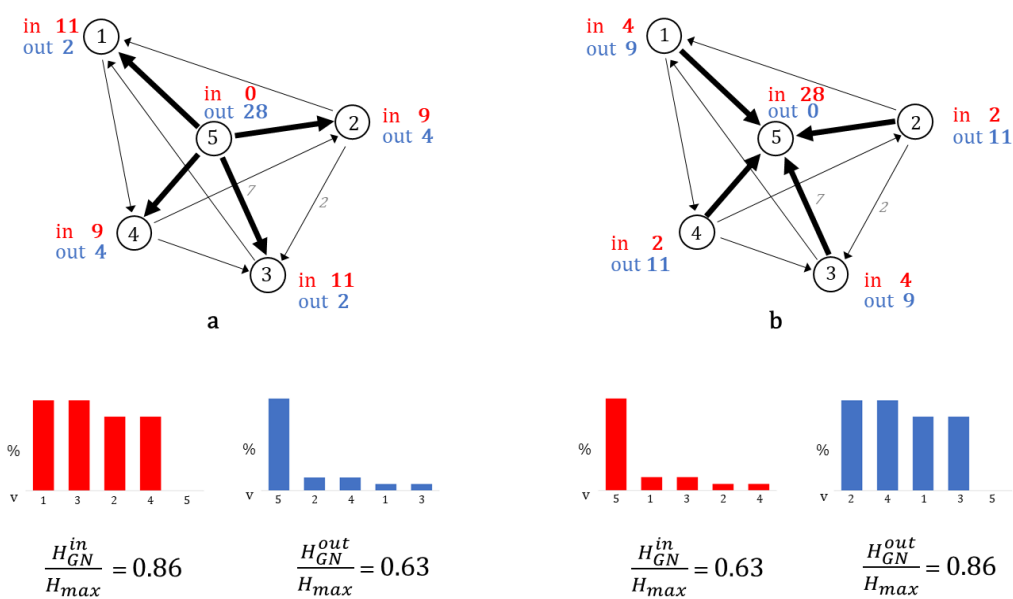}}
\caption{Global nodal entropy. Example of two different commuting networks with the corresponding distribution of flows among nodes, in-flows in red and out-flows in blue.}
\label{Gl_nodes}
\end{figure}

Both measures, reveal topological patterns of the network according to commuting origin or destination, providing information about the concentration of flows. In the case when one node concentrates most of the flows, the system will exhibit a skewed probability distribution, indicating that if a location is taken at random, it will most likely correspond to that node. This reduced uncertainty of knowing where an individual randomly selected might go to work (or come from), is represented through a lower entropy. If on the other hand, there is a similar probability distribution of flows across nodes, such that the system has no node dominating over others, the uncertainty of ascertain the work or home location of an individual will be higher and hence the entropy will also be higher. The latter is maximal when there is equiprobability across space. 
Figure \ref{Gl_nodes} exemplifies the centralisation of system \textbf{b} with respect to in-commuters, concentrating most of the flows in $v_{5}$, while for the same configuration and different directions, system \textbf{a} does not concentrates job, attaining hence a higher in-flow entropy than \textbf{b}. The inverse occurs for the out-flows.

In general, commuting destinations tend to be more highly concentrated than the commuting origins. This is because employment opportunities tend to cluster in few locations. In those cases the out-entropy will be higher than the in-entropy. However, this is not always the case, and exploring whether the origins or destinations of the commuter flows are more or less diverse, by looking at whether $H_{GN}^{out}>H_{GN}^{in}$ or $H_{GN}^{in}>H_{GN}^{out}$, can give a better understanding of the urban system. Figure \ref{Gl_nodes} presents a strong case of monocentricity with respect to jobs in system \textbf{b}, where $H_{GN}^{out}>H_{GN}^{in}$, and the odd case from which a single location provides most workers for several different locations in system \textbf{a}, with $H_{GN}^{in}>H_{GN}^{out}$.

To normalise the results and make them comparable between systems of different sizes (different number of nodes), we look for the maximum possible value for each system, which is $H_{Tot}$= $\log(n)$, where $n$ is the total number of nodes in the system. The normalised entropies can be written as:  $H_{GN}^{out}/H_{Tot}$ and  $H_{GN}^{in}/H_{Tot}$.

\subsubsection*{Commuter trips distribution}

In the previous section we characterised the origins and destinations of commuting flows according to their diversity, and considered how such an approach can give insights into the polycentricity of cities. Let us now look at the trips that are being generated, and measure the diversity of the flows along the links of the commuting network.

The concept of diversity is associated in this case, with the dispersion of commuter trips in the territory taking into account their intensity and density. Therefore, both the variety of areas that are connected to each other, as well as the equivalence between the intensity of their interactions are considered. A system will be more diverse if there are many combinations of origin-destination pairs, and if the amount of flows between these pairs is evenly distributed. 

The framework presented here is relevant to inform infrastructure planning, given that the provision of transport infrastructure is intertwined with the spatial distribution of flows. The more disperse the pattern of origins and destinations in the territory are, the more challenging is the planning of the physical transport structure that allows these trips to occur more efficiently \cite{Lowe1998}. In addition, such a framework also allows us to identify functional dependencies between urban units within the system. If the relationships are scattered it means that the operation of the system relies on various labour and economic relationships between its different components. The opposite occurs with the existence of dominant flows where most of the trips occur between few pairs of urban areas, and the overall system is constrained to these specific relationships.

Let us introduce global entropy at link level, as a measure of flow diversity, considering every OD pair in the system. We normalise the measure with respect to its maximum, so that comparisons with other systems can be made. In this case, we need a \textit{joint entropy} encompassing the uncertainty associated with both origin and destination, through the link probability. Such a measure can be interpreted as an average diversity of the system as a whole \cite{Sole2004}. The entropy of trips can be defined as:

\begin{itemize}
\item Global entropy at link level:

\begin{equation}\label{eq:(2)}
 H_{GL}= -\sum_{\forall i} \sum_{\forall j}  p_{ij}  \log p_{ij}
\end{equation} 
where $p_{ij}$ is the probability that a commuting flow from $v_{i}$ to $v_{j}$ occurs in the system, hence $p_{ij} = \frac{w_{ij}}{\sum_{i}\sum_{j} w_{ij}}$, where $w_{ij}$ is the number of trips from $v_{i}$ to $v_{j}$.

\end{itemize}

\begin{figure}[h!]
\centering
\makebox[\textwidth][c]{\includegraphics[width=1.15\textwidth]{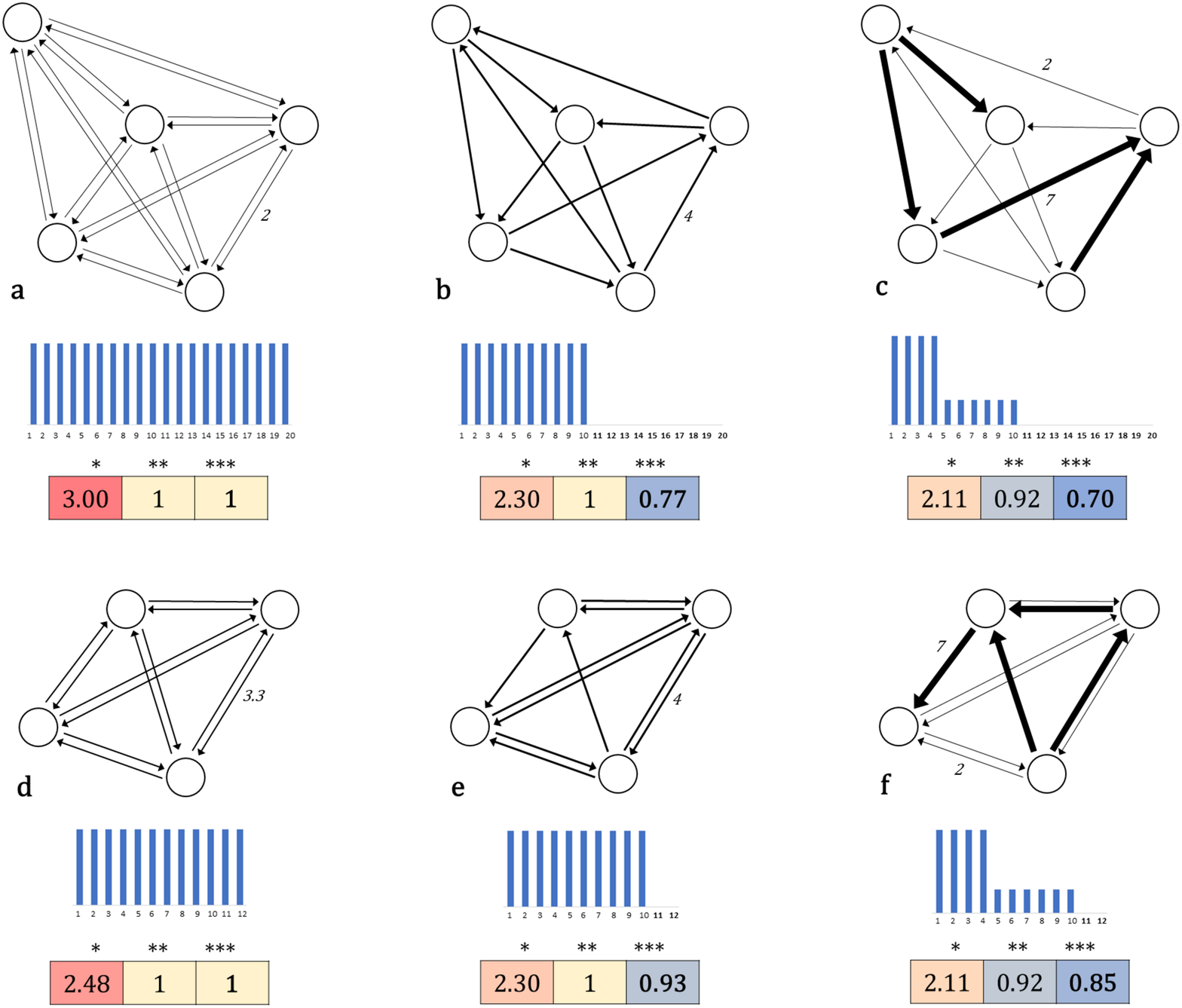}}
\caption{Global links entropy. Example of commuting networks with the distribution of link weights and the corresponding general entropy and normalised values: (*) = $H_{GL}$, (**)= $H_{GL}$/$H_{Tot}$, and (***)= $H_{GL}$/$H_{max}$. Values are coloured red to blue, from higher to lower, to facilitate comparison.}
\label{GL_links}
\end{figure}

$H_{GL}$ takes higher values when flow weights are evenly distributed, so every commuting flow is equally relevant in the commuting network. Conversely, if only few OD links contain the large majority of commuting flows, the diversity of the system is low. Then, the dominance of some flows in the network reduces the entropy $H_{GL}$. This is clear when looking at networks \textbf{b} and \textbf{c} in Figure \ref{GL_links}. With the same total flow count and the same amount of links, but with a different distribution of flows among them, the global entropy at link level is higher in \textbf{b} than in \textbf{c}. In \textbf{b} flows are evenly distributed, while in \textbf{c} certain links have a much higher density than in the rest of the system.

In general, entropy values tend to be higher when the number of elements in the system increases, so the more the links $m$  or nodes $n$  in the network, the higher the entropy values. In Figure \ref{GL_links}, we can confirm this by looking at networks \textbf{a} and \textbf{d}. In both systems flows are evenly distributed across links, however entropy (*) is higher in \textbf{a} than \textbf{d} because $m_{\textbf{a}} > m_{\textbf{d}}$. This means that the comparison of different systems is not a straightforward task. To address this issue we need to normalise our measures of entropy. A common way of addressing normalisation when studying entropy is by looking at its maximum value, which occurs when all elements are equally abundant. Then, normalisation is done by dividing the entropy value by the entropy of the total number of elements present in the system. Accordingly, diversity of link weights in a network with $m$ links would be normalised by $H_{Tot}= \log (m)$: $H^{**}=H_{GL}/H_{Tot}$ \cite{Pielou1966}.

A more suitable normalisation when comparing diversity of commuting flows between system of cities should consider a notion of density. Network density in this context, is understood as the ratio of the total number of links  in the network to the number of links in its theoretical fully connected network \cite{Green2007}. The maximum possible number of links in a graph is given by $n(n-1)$, $n$ being the number of vertices in the graph. The proposed normalisation in this work is then by $H_{max}= \log(n(n-1))$, leading to $H^{***}=H_{GL}/H_{max}$. This process takes into account the diversity of flows in a existing level of interaction in a system, in comparison to its own maximum potential of connectivity. The latter, allows a meaningful comparison of flow diversity between systems of different sizes (different number of commuting OD pairs or different number of cities).

A comparison between both forms of normalisation could be easily done by looking at examples \textbf{b} and \textbf{e} in Figure \ref{GL_links}. Taking the common form of normalisation: $H^{**}(\textbf{b}) = H^{**}(\textbf{e}) = 1$, giving the maximum value to both systems which have the same amount of existing links ($m = 10$) and even distribution of weights. By this, we could conclude that both systems are equally diverse. However, when taking the proposed normalisation (***) the results for each systems are $1> H^{***}(\textbf{e}) > H^{***}(\textbf{b})$. As this measure considers density, in this case none of the systems meet the maximum entropy value of 1, because they do not present the maximum possible number of OD pairs according to their own potential, as in the cases of \textbf{a} and \textbf{d}. We observe that the density of links in \textbf{e} is closer to its maximum potential, in comparison to \textbf{b}.

\subsection{Local diversity}

In addition to identifying general characteristics of the network, we are also interested in understanding the role of individual locations. Local measures serve this purpose, and local diversity can be thought of as a sub-graph entropy of the node in question, where every in or out link directly connected to it is taken into account. This measure considers the intensity and density of the trips that are released (outflow) or attracted (inflow) by each unit (node). The intensity of flows informs whether the relationships are distributed in a scattered or polarised manner. Density, on the other hand, looks at the variety of urban areas with which the unit in question interacts. 

Identifying important actors in the distribution of flows is important to be able to construct decentralised solutions. These are favoured to increase the resilience of the network. Decentralisation can be achieved by diversifying the dependence between nodes. Looking at the specific case of commuting networks, the distribution of inflows is determined by the areas of provision of labour for the internal employment market. On the other hand, the distribution of workforce outflows accounts for dependencies between residents in a certain area, and the provision of jobs in other locations. Within this proposed framework, areas of similar interaction and dependency patterns can be identified, from which a categorisation of cities can be constructed to inform planning decisions.  For in- and out-commuting scenarios, local entropy is defined as:

\begin{itemize}
\item Local in-flow entropy:

\begin{equation}
 H_{L}^{in}= -\sum_{\forall i}  p_{(i|j)} \log p_{i|j} =  -\sum_{\forall i}  \frac{p_{ij}}{p_{j}}  \log  \frac{p_{ij}}{p_{j}}
\end{equation} 

\item Local out-flow entropy:
\begin{equation}
 H_{L}^{out}= -\sum_{\forall j}  p_{j|i}  \log  p_{j|i} = -\sum_{\forall j}  \frac{p_{ij}}{p_{i}}  \log  \frac{p_{ij}}{p_{i}}
\end{equation} 
where $p_{j} = \frac{\sum_{i} w_{ij}}{\sum_{i}\sum_{j} w_{ij}}$ represents the sum of every flow arriving at $v_{j}$ divided by the total flow of the system, and $p_{i} = \frac{\sum_{j} w_{ij}}{\sum_{i}\sum_{j} w_{ij}}$ refers to the sum of every flow leaving $v_{i}$ divided by the total flow of the system. $p_{(i|j)}$ and $p_{(j|i)}$ represent the probability that a flow within the system is received or sent by a specific node respectively.

\end{itemize}

\begin{figure}[h!]
\centering
\makebox[\textwidth][c]{\includegraphics[width=1.15\textwidth]{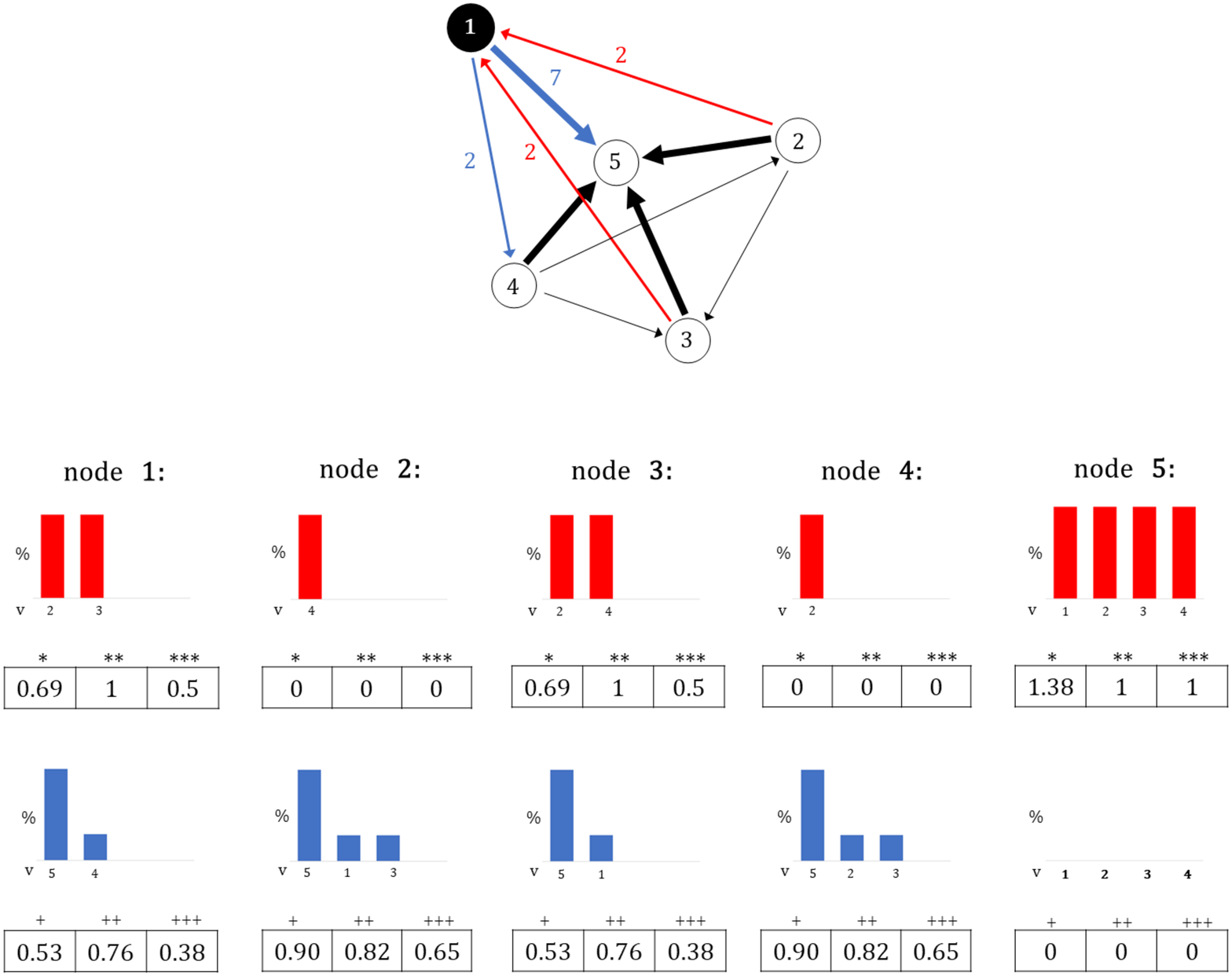}}
\caption{Local nodes entropy. A commuting network where the sub-graph at node $v_{1}$ is highlighted with in-flows in red, and out-flows in blue. Diagrams for in-flows distribution at every node in red, with entropy and normalised values corresponding to: (*)= $H_{L}^{in}$, (**)= $H_{L}^{in} /H_{\text{deg}}^{in}$, and (***)=$H_{L}^{in} /H_{max}$. In blue the out-commuting distribution, with: ($^+$)= $H_{L}^{out}$, ($^{++}$)= $H_{L}^{out} /H_{\text{deg}}^{out}$, and ($^{+++}$)= $H_{L}^{out} /H_{max}$.}
\label{Lc_links}
\end{figure}

These measures give information about the node diversity in terms of the flows that are sent or received by its direct neighbours (one-hop neighbours of the target node). In this case, entropy functions are applied to the distribution of flow weights to or from a given node. Thus, the dominance of an origin-destination pair at a given node reduces the entropy, while an equal distribution of flows results in higher values of entropy. 
Figure \ref{Lc_links} shows the example of a node $v_{1}$ whose in-links have equal weights, while the out-links are dominated by flows commuting to $v_{5}$, leading to $H_{L}^{in} > H_{L}^{out}$. An entropy equal to zero occurs when there is no link arriving or departing from a node (e.g. out-commuting from $v_{5}$ in Figure \ref{Lc_links}). But, this will also be the case when there is only one link connected to the node, since there will be no uncertainty (e.g. in-commuting from $v_{2}$ or $v_{4}$ in Figure \ref{Lc_links}).

Normalisation can be achieved by dividing by the maximum value of entropy given the in or out degree of the node:  $H_{\text{deg}}^{in}=\log(\text{deg}_{in}(v_{j}))$ and $H_{\text{deg}}^{out}= \log(\text{deg}_{out}(v_{i}))$. Another approach for normalising local nodal entropy could be done by looking at its maximum possible value, which in this case would be given by $H_{max}= \log(n-1)$. As explained previously for equation (\ref{eq:(2)}), the proposed framework takes into consideration the network density and the maximum potential of connectivity of nodes in a network. The normalised diversity measures are: $H_{L}^{in} / H_{max} $ and $H_{L}^{out}/ H_{max}$. The relevance of the normalisation is illustrated in Figure \ref{Lc_links}. 
With the first normalisation (**), looking at the in-flows, both nodes $v_{5}$ and $v_1$ present the highest value, since flows are equally distributed among the existing links. In the second normalisation (***), node $v_{5}$ has a higher value since it receives flows from every possible node in the network, while $v_{1}$ only receives flows from half of the potential origins.

If we want to describe the system based on its local relationships, we can compute the average among every local entropy. The  following equations measure the weighted mean where every local measure in the system is considered based on its different probability of occurrence.  In Information Theory this measure is known as \textit{conditional entropy}, and it quantifies the uncertainty about a variable when another variable is known \cite{Wilhelm2007}. The first measure corresponds to the uncertainty of in-commuting when destination is known, the second one corresponds to the uncertainty of out-commuting when origin is known: 

\begin{itemize}
\item Average local in-flow entropy:
\begin{equation}
 H_{L\mu}^{in}= \sum_{\forall j} p_{j} (H_{L}^{in})_{j}= -\sum_{\forall i} \sum_{\forall j} p_{ij}  \log  p_{(i|j)} = -\sum_{\forall i} \sum_{\forall j}  p_{ij}   \log \frac{p_{ij}}{p_{j}}
\end{equation} 

\item Average local out-flow entropy:
\begin{equation}
 H_{L\mu}^{out}= \sum_{\forall i} p_{i} (H_{L}^{out})_{i}= -\sum_{\forall i} \sum_{\forall j} p_{ij}  \log  p_{(j|i)} = -\sum_{\forall i} \sum_{\forall j} \; p_{ij} \log \frac{p_{ij}}{p_{i}}
\end{equation} 
\end{itemize}

Note that the average is obtained by considering all possible values of $i$ or $j$ given by each probability of occurrence $p_{i}$ or $p_{j}$. Normalisation of these measures could be done by dividing the results by $\log(n-1)$, in the same way as for equation (5) and (6), allowing us to compare diversity in different systems at local scales in terms of their own maximum potential of connectivity.

\subsection{Multiple measures analysis}

\begin{figure}[h!]
\centering
\makebox[\textwidth][c]{\includegraphics[width=1.15\textwidth]{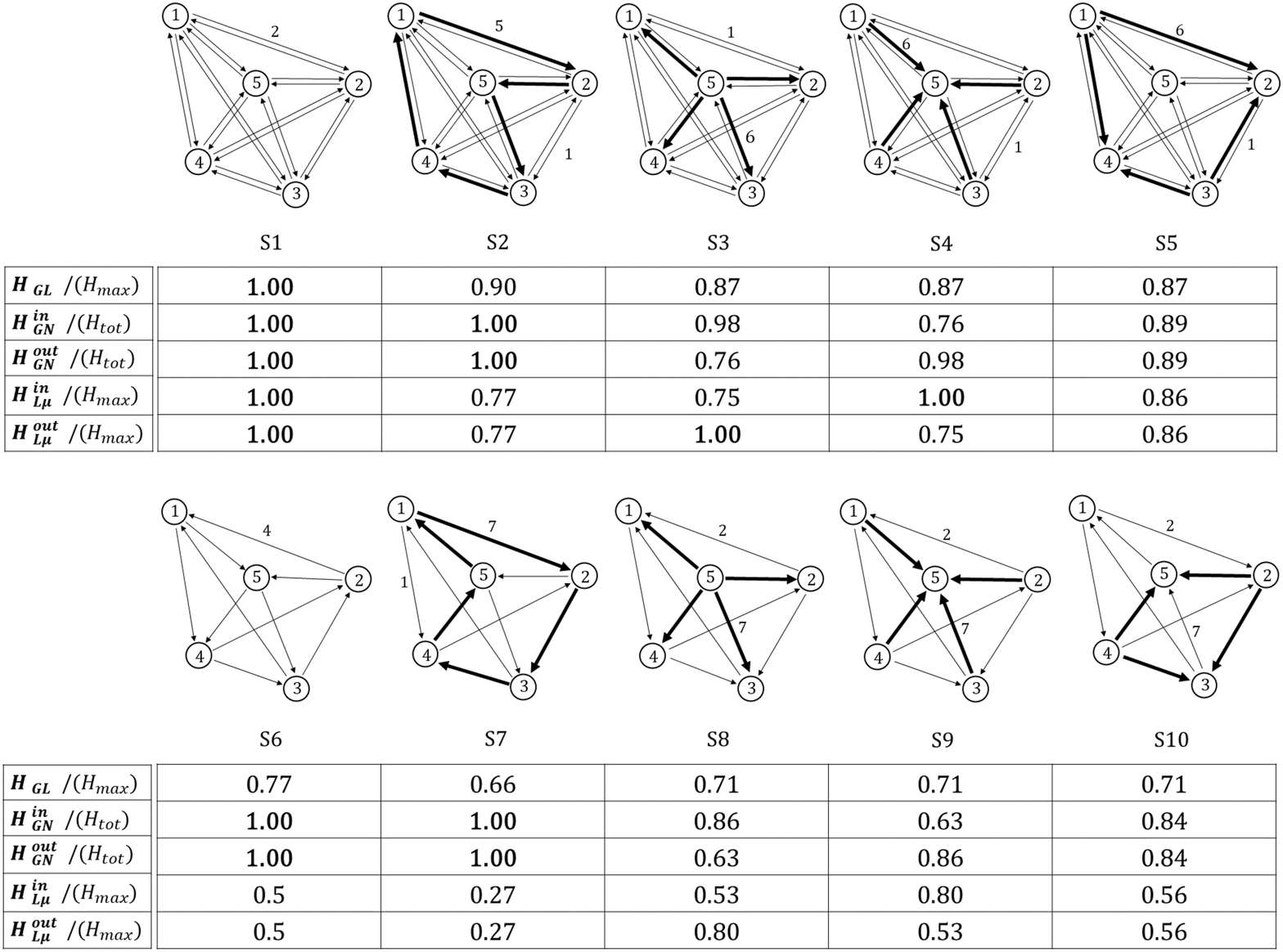}}
\caption{Global and local measures across different commuting networks. Values correspond to the normalised entropy. The total flow count is the same across all systems.}
\label{all_NTs}
\end{figure}

In this paper we have explored some entropy functions at the global and local level for directed networks, aiming at capturing different relationships between system components. When looking at Figure \ref{all_NTs}, where all revised measures are computed across 10 toy models with different flow distributions but same topology, we can observe the high variability of outputs. This tells us how important it is to choose the proper measure to describe the pattern of interest. A single measure will not be able to capture the complex structure of flow diversity in the system. This is clearly shown in system $S7$ in Figure \ref{all_NTs}, where values of entropy vary between 0.27, the minimum value in the whole table, and the maximum 1. 

$H_{GN}^{out}$ and $H_{GN}^{in}$ are able to capture concentration of flows which could potentially inform about the presence of predominant centres. If we look at the first row of networks in Figure \ref{all_NTs}, $H_{GN}^{in}$ results can be sorted as $S1=S2 > S3 > S5 > S4$. In the first two systems, flows are equally distributed among nodes, having the maximum diversity. The network S5 presents polycentricity, where flows are mainly clustered in two nodes. While network S4 presents monocentricity, with a concentration of flows at one destination. This system has a lower $H_{GN}^{in}$  value. On the other hand, in order to differentiate patterns between S1 and S2 we need another measure. By looking at the $H_{GL}$ values, we can observe that $S1 > S2$. This means that while in both systems nodes are equally relevant, the relationships between nodes are different. In S2 the measure is able to capture that some of the origin-destination pairs dominate over others. 

\begin{figure}[h!]
\centering
\makebox[\textwidth][c]{\includegraphics[width=1.15\textwidth]{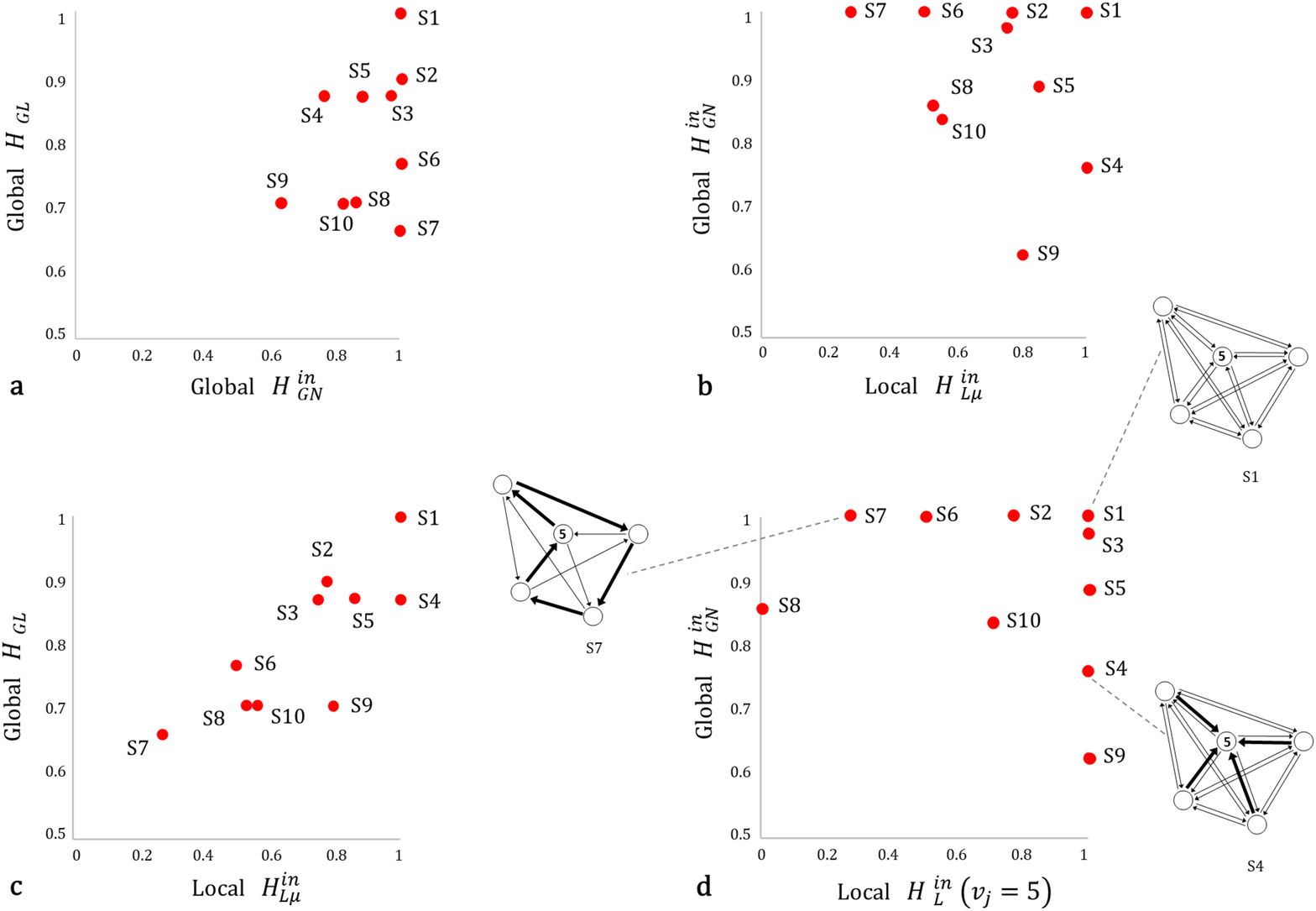}}
\caption{Comparison between local and global normalised measures of networks in Figure \ref{all_NTs}. Figure \textbf{a} compares both global measures, \textbf{b} and \textbf{c} compare global with average local entropy, and finally \textbf{d} compares global with local entropy at node $v_{5}$. System S1 is a fully connected network with evenly distributed flows with the maximum entropy in every measure, while other systems present varying patterns across measures depending on their structural organisation. As an example we can see that in \textbf{d} node $v_{5}$ in S1 has high entropy within a globally diverse system, and in contrast node $v_{5}$ in S4 has high local entropy within a system which is centrally organised. Although  individual node interactions are equally distributed among neighbours in S4, node $v_{5}$ has a preponderant role globally. In S7 node $v_{5}$ has low local entropy but is part of a fully diverse system. }
\label{plots}
\end{figure}

Local measures for commuting networks are more commonly used to better understand the dynamics occurring at place level. In a regional system for example, local diversity allows us to capture how heterogeneous is the interconnection of a city with other cities in the system. In this context, the average of all local measures among cities, is expected to reflect the dynamics occurring among all interconnections between the whole system components. However, we believe that this is not a straightforward assumption, and it is necessary to question to what extent local measures could capture the global diversity of mobility, and how both local and global relates to each other. In Figure \ref{plots}, we explore those relationships between local and global measures of diversity across toy models presented in Figure \ref{all_NTs}. 
Figure \ref{plots}-\textbf{b} shows the relationship between global $H_{GN}^{in}$ and local $H_{L\mu}^{in}$  in-commuting measures. In general we can see that both reflect different aspects of diversity in the system, and they do not present any obvious correlation. For example, system S4 has maximum local diversity but a relative low global diversity. 

A city is not fully characterised by the relationships established with its immediate neighbours, its role within the wider context and dynamics of the region, country or trade network it belongs to, play an important role in its characterisation. For example, a city with a high local mobility diversity within a diverse region, will not have the same role as a city with the same local diversity but within a non-diverse regional system. In Figure \ref{plots}-\textbf{d}, we compute local in-commuting node entropy $H_{L}^{in}$ for node $v_{5}$ in each system in Figure \ref{all_NTs}, and plot it against the global entropy $H_{GN}^{in}$ of its correspondent network. This shows that the node $v_{5}$ in networks S1, S3, S4, S5 and S9 is fully diverse locally, with the same entropy value, but is part of systems that behave completely differently globally. Taking S1 and S4, as opposite examples, we can clearly see that in network S4, the node $v_{5}$ has a dominant role, functioning as a centre of destinations.  Conversely in S1, node $v_{5}$ has the same role as every other component within the overall network.

\section{Discussion}

Through this work we have shown that the application of entropy theory in the analysis of commuting networks provides relevant information on the distribution of flows in the territory. We explored measures of entropy on global and local scales, as well as on the different constituent elements of the network of flows, links and nodes. Each measure proved relevant in capturing distinct aspects of the spatial interaction patterns. 

Link entropy focuses on the interactions between pairs of urban areas, based on the distribution of origin-destination trips. Nodal entropy on the other hand, gives us information on the concentration or dispersion of flows among urban centres. The local analysis examines the relationships between labour supply and demand that a specific area establishes with its most direct context. When extending the analysis to the larger scale, each of the interactions occurring in the system are considered. All the constituent elements, whether they are or not directly or intensely connected to each other, influence the whole-system's entropy.

The latter is particularly useful if for example, we want to analyse the resilience of commuting networks based on the diversity of the structure given by connectivity. Systems can face direct or indirect changes that occur at different levels. Local entropy will change when endogenous changes in the local labour market alter the structure of relationships of an urban area. On the other hand, changes in global entropy can account for exogenous changes that occur in other local systems. These changes can end up affecting the global structure to a greater or lesser extent, and therefore indirectly modify the structure of relationships between all constituent parts.

It is worth mentioning that with this analysis we are not studying the optimum degree of diversity in the system, nor are we arguing that the maximum possible entropy should be pursued. The functioning of urban systems must be flexible enough to adapt to changes and at the same time efficient enough to optimise resources. As Cabral et al. (2013) argue, if the system falls short of a minimum entropy, the system will be very centralised and therefore vulnerable to changes, while if it exceeds a certain degree of entropy, the system will not be dealing efficiently with resources \cite{Cabral}. The distribution of workplaces and housing requires a certain degree of concentration to benefit from specialisation and proximity, but at the same time a degree of diversity and dispersion would increase the capacity for resilience and adaptability \cite{Goetz2010}. Consequently, the interpretation of the results of the different entropy measures presented in this paper must be made based on the specific criteria of the system under study.

The different measures of entropy presented here contribute to advancing our understanding of the complexity of spatial flows, to inform policy development and take strategic planning actions. By analysing the entropy relative to a maximum possible number of interactions of the system, instead of the given or existing ones, it is possible to compare the system with itself in terms of its maximum capacity.  We believe that this form of normalisation presented in this paper facilitates the study of systems based on their own potentialities, offering a different perspective for planning. 

This introductory work contributes to the understanding of real commuting networks across many different scales of organisation, in addition to providing a framework to better understand the interplay between transport infrastructure and the layout of economic opportunities in cities.


\section*{Acknowledgements}

VM thanks The National Research and Development Agency (ANID) for the financial support provided as a PhD scholarship "Becas Chile".

\section*{Author contributions statement}
VM developed the theoretical framework, designed the experiments and conducted the analysis. CM and EA contributed to the theoretical framework . VM wrote the paper, and all authors contributed to the structure and fine tuning of the manuscript.

\end{document}